\documentstyle[12pt]{article}

\def\hybrid{\topmargin -20pt  \oddsidemargin 0pt
      \headheight 0pt   \headsep 0pt
      \textwidth 6.25in % A4 paper
      \textheight 9.5in % A4 paper
      \marginparwidth .875in
      \parskip 5pt plus 1pt   \jot = 1.5ex}

\hybrid

\begin{document}
%\titlepage
\def\x{\times}
\def\beq{\begin{equation}}
\def\eeq{\end{equation}}
\def\beqa{\begin{eqnarray}}
\def\eeqa{\end{eqnarray}}
\def\L{{\cal L}}
\def\M{{\cal M}}
\def\K{{\cal K}}
\def\N{{\bf N}}
\def\.{{\bullet}}

\sloppy
\newcommand{\be}{\begin{equation}}
\newcommand{\eq}{\end{equation}}
\newcommand{\ov}{\overline}
\newcommand{\un}{\underline}
\newcommand{\p}{\partial}
\newcommand{\la}{\langle}
\newcommand{\ra}{\rangle}
\newcommand{\bl}{\boldmath}
\newcommand{\ds}{\displaystyle}
\newcommand{\nl}{\newline}
\newcommand{\th}{\theta} 

%\textwidth14.5cm
%\textheight23.0cm
%\oddsidemargin0.5cm
%\topmargin-1.4cm

%\addtocounter{section}{1}

\renewcommand{\thesection}{\arabic{section}}
\renewcommand{\theequation}{\thesection.\arabic{equation}}

%\setcounter{section}{1}
%\addtocounter{section}{1}
\parindent1em
%\newcommand{\resetcounter}{\setcounter{equation}{0}}     
% set counter to zero
\vfill
\eject

\begin{titlepage}
\begin{center}
\hfill HUB-EP-98/16\\
\hfill {\tt hep-th/9802202}\\

\vskip .7in

{\bf  ON VECTOR BUNDLES AND CHIRAL MATTER IN N=1 \vskip 0.1cm
      HETEROTIC COMPACTIFICATIONS}

\vskip .5in

Bj\"orn Andreas \footnote{email:
andreas@qft3.physik.hu-berlin.de}
 \\
\vskip 0.3cm
{\em Humboldt-Universit\"at zu Berlin,\\
Institut f\"ur Physik, 
D-10115 Berlin, Germany}

\vskip .1in

\end{center}

%\begin{center} {\bf ABSTRACT } \end{center}
\begin{quotation}\noindent

In this note we derive the net number of 
generations of chiral fermions in heterotic
string compactifications on Calabi-Yau threefolds
with certain $SU(n)$ vector bundles, for n odd, using the parabolic 
approach for bundles. We compare our results with the
spectral cover construction for bundles and make a
comment on the net number interpretation in F-theory.

\end{quotation}
\end{titlepage}
\vfill
\eject

%------------------------------------------------------------------------------
\newpage
%------------------------------------------------------------------------------
\section{Introduction}
%------------------------------------------------------------------------------
Recent progress has been made in our understanding of heterotic string 
compactification on Calabi-Yau threefold $Z$ with vector bundle $V$ embedded
in $E_8\times E_8$ [\ref{FMW}] as well as in its expected dual F-theory 
compactification 
on Calabi-Yau fourfold $X$ [\ref{V},\ref{MV},\ref{BJPS}] which both lead to 
$N=1$ supersymmetry in four 
dimensions. Part of the progress was the appearance (independent of any 
duality consideration) of non-perturbative heterotic five-branes $n_5$ 
which are necessary for a consistent heterotic compactification [\ref{FMW}]. 
On the F-theory side, the necessity of turning on a number   
$n_3$ of space-time filling three-branes for tadpole cancellation [\ref{SVW}]
which match the heterotic five-branes [\ref{FMW},\ref{AC}] was established. 
Further, it was shown that this matching can be refined due to the 
occurrence of some discrete data on the heterotic -and F-theory 
side [\ref{CD}]. In particular, it is expected that the associated moduli 
spaces of the two theories are isomorphic 
[\ref{FMW},\ref{BJPS},\ref{CD}-\ref{AM}].

Apart form the matching of moduli in both theories, which lead to massless
gauge neutral chiral matter, one should also understand the part of the 4D
heterotic spectrum which corresponds to massless under the gauge group 
charged matter as well as chiral matter in terms of geometry on the 
F-theory side. Progress in this direction was made in 
[\ref{BJPS},\ref{BCGJL}-\ref{ler}].

Besides the spectrum matching, further checks of het/F-theory duality involve
the comparision of $N=1$ effective interactions which are determined by 
holomorphic quantities, i.e. superpotentials and gauge kinetic functions
[\ref{sup1}-\ref{sup4}].

Let us be more concrete! The heterotic/F-theory duality picture was
established by considering $Z$ as an elliptic fibration $\pi:Z\rightarrow B_2$
with a section $\sigma$, where $B_2$ is a two-fold base; $Z$ can be represented
as a smooth Weierstrass model. $X$ was considered as being elliptically
fibered over a threefold base $B_3$, which is rationally ruled, i.e. 
a fibration $B_3\rightarrow B_2$ with $P^1$ fibers exists because we have to
assume that 
the fourfold is a $K3$ fibration over the twofold base $B_2$ in order
to extend adiabatically the 8D duality between the heterotic string on $T^2$
and F-theory on $K3$ over the base $B_2$.

In addition to the specification of $Z$, we have to specify a stable
vector bundle 
over $Z$, which breaks part of the $E_8\times E_8$ gauge symmetry. We will
consider a $G=SU(n)$ vector bundle which determines a rank $n$ complex 
vector bundle $V$ of trivial determinant. Actually, there are
three methods for constructing stable vector bundles over
elliptic fibrations: the parabolic, the spectral cover and the construction
via del Pezzo surfaces, which are explained and developed in [\ref{FMW}] and 
[\ref{D2},\ref{FMW2},\ref{FMW3}].
We will adopt the parabolic construction here, since it allows us
to easily compute Chern-classes, however we will compare our results to 
those obtained from the spectral cover construction. In particular it will
be shown that both approaches agree for a certain choice of the 
twisting line bundle on the spectral cover. 

In the parabolic approach [\ref{FMW}] considered a 
component of the moduli space of 
$SU(n)$ bundles (understood as rank $n$ complex vector bundles) which are 
$\tau$-invariant, i.e. the involution of the elliptic fibration 
$Z\rightarrow B$ lifts up to $V$. This condition 
could be implemented for $SU(n)$ bundles with $n$ even. 

In the spectral cover approach [\ref{FMW}] considered  $V=SU(n)$ bundles 
with $n$ arbitrary. These bundles possess an additional 
degree of freedom since one has the possibility to 'twist' with a line 
bundle ${\cal N}$ on the spectral cover $C$, which leads to a multi-component 
structure of the moduli space of such bundles. Further, they did not require 
any $\tau$-invariance of $V$ in the construction, moreover, they 
found $\tau$-invariance for a certain class of these bundles which
have no additional twists. In particular, they found the same modulo conditions
(see below) as in the parabolic approach, in order to get $\tau$-invariance. 
Further, their $\tau$-invariant bundles have vanishing third Chern-class.

In {\it section 2.1.} we will first review, for our purposes, some 
facts of the 
parabolic construction for $V$ an $SU(n)$ vector bundle, then we will recall
what is known for $n$-even which was discussed in [\ref{FMW}] and we
will extend this to the case if $n$ is odd. In contrast 
to [\ref{FMW}] we do not 
focus on a $\tau$-invariant point in the moduli space of $SU(n)$ bundles, 
this allows us to determine the net number $N_{gen}$ of generations of 
chiral fermions, i.e.
($\sharp$ generations $-$ $\sharp$ antigenerations)
in the observable sector of the 4D unbroken gauge group. In contrast to
the bundles at the $\tau$-invariant point, which have $n$ even and 
a modulo 2 condition for $\eta$, our bundles have $n$ odd and 
$\pi_*(c_2(V))=\eta$ divisible by $n$. 
Then we will proceed in {\it section 2.2.} and compare them with the spectral 
cover 
construction of $V$ and in {\it section 2.3.} conclude with a comment
on certain discrete data in F-theory which turn out to be related to 
$N_{gen}$.
%------------------------------------------------------------------------------
\section{Bundle Construction}
%------------------------------------------------------------------------------
\subsection{Parabolic Approach}
%------------------------------------------------------------------------------

In the parabolic approach we start with an unstable
bundle on a single elliptic curve $E$ which is given by{\footnote{
here $W_k$ can be defined inductively as a unique non-split extension
$0\rightarrow {\cal O}\rightarrow W_{k+1} \rightarrow W_{k}\rightarrow 0$
(c.f.[\ref{FMW}])}}
\beqa
V=W_k\oplus W_{n-k}^*
\eeqa
this has the property that it can be deformed by an element $\alpha\in H^1(E,
W^*_k\otimes W^*_{n-k})$ to a (semi) stable bundle $V'$ over $E$ which fits in
the exact sequence 
\beqa
0\rightarrow W^*_{n-k}\rightarrow V'\rightarrow W_k\rightarrow 0
\eeqa
Now, to get a global version of this construction, we are interested in 
an unstable bundle over $Z$, which 
reduces on every fibre of $\pi:Z\rightarrow B$ to the unstable 
bundle over $E$ and which can be deformed to a stable bundle over
$Z$. Since the basic bulding blocks were on each fibre $W_k$ with 
$W_1={\cal O}(p)$ we have to replace them by their global versions.  
So, one replaces global $W_1={\cal O}(p)$ by $W_1={\cal O}(\sigma)$ 
and $W_k$ can be defined inductively by an exact sequence
\beqa
0\rightarrow {\L}^{n-1}\rightarrow W_k \rightarrow W_{k-1}\rightarrow 0
\eeqa
with ${\cal L}=K_B^{-1}$.
Further, one has globally the possibility to twist by additional data coming
from the base $B$ and so one can write [\ref{FMW}] for $V=SU(n)$ 
\beqa
V=W_k\otimes{\cal M}\oplus W^*_{n-k}\otimes{\cal M}^{\prime}.
\eeqa
Note that the unstable bundle can be deformed to a stable one by an element
in $H^0(B,R^1\pi_*(ad(V)))$ since the Leray spectral sequence degenerates
to an exact sequence [\ref{FMW}]. Further, note that we can use the 
unstable bundle for the computation
of characteristic classes, because the topology
of the bundle is invariant under deformations.  
%------------------------------------------------------------------------------

%------------------------------------------------------------------------------
Now, let us start with an unstable $G=SU(n)$ bundle $V$
where ${\cal M}, {\cal M}^{\prime}$ are line bundles over $B$ which are 
constrained so that $V$ has trivial determinant, i.e. ${\cal M}^k\otimes({\cal M}^{\prime})^{n-k}\otimes{\L}^{-\frac{1}{2}(n-1)(n-2k)}\cong{\cal O}$.
Further, $W_k$ and $W^*_{n-k}$ are defined as
\beqa
W_k=\bigoplus_{a=0}^{k-1}{\L}^a, \ \ \ \
W^*_{n-k}=\bigoplus_{b=0}^{n-k-1}{\L}^{-b}
\eeqa
where we have set ${\L}^0={\cal O}(\sigma)$ and 
${\L}^{-0}={\cal O}(\sigma)^{-1}$ with $\L$ being a line bundle on $B$. 
The total Chern class of $V$ can be written as
\beqa
c(V)=\prod_{a=0}^{k-1}(1+c_1(\L^a)+c_1(\M))
     \prod_{b=0}^{n-k-1}(1+c_1(\L^{-b})+c_1(\M^{\prime})).
\label{total}
\eeqa
Now we will discuss the following two cases: $n$ is even and $n$ is odd.
First let us review the case that $n$ is even which was considered
in [\ref{FMW}], then we will extend to the case $n$ is odd and compare both
cases with the spectral cover construction of $V$.\\ \\  
{\underline{$n$ even}}\\ \\
In this case, one can choose $k=\frac{n}{2}$ which
restricts one to take ${\cal M}^{\prime}={\cal M}^{-1}$ in order to obey
the trivial determinant of $V$. The advantage of taking $k=\frac{n}{2}$ is that
the condition of $\tau$ invariance of $V$ is easily implemented.  
Note that $\tau$ operates on $V$ as $\tau^*V=V^*$, i.e. $k\rightarrow n-k$.
Now, the expansion of the total Chern class of $V$ 
immediatly leads to $c_1(V)=0$ and $c_3(V)=0$. Further setting 
$\sigma=c_1({\cal O}(\sigma))$ and $\eta=-2c_1(\M)+c_1(\L)$
respectively using the fact that $\sigma^2=-\sigma c_1(\L)$, one obtains  
for the second Chern class 
\beqa
c_2(V)=\eta\sigma-\frac{1}{24}c_1(\L)^2(n^3-n)-\frac{n}{8}\eta(\eta-nc_1(\L)).
\eeqa    
%\beqa
%c_2(V)=-2\sigma c_1(\M)-\sigma^2+\left(\frac{-n^3+3n^2-2n}{24}
%\right)c_1(\L)^2+
%\left(\frac{2n-n^2}{4}\right)c_1(\M)c_1(\L)\\-\frac{n}{2}c_1(\M)^2
%\nonumber
%\eeqa
Moreover, from $c_1(\M)=-\frac{1}{2}(\eta-c_1(\L))$ one gets the congruence 
relation for $\eta$:
\beqa
\eta\equiv c_1(\L) \ \ (mod\ \ 2)
\eeqa
Thus $c_2(V)$ is uniquely determined in terms of $\eta$ and the elliptic
Calabi-Yau manifold $Z$. In particular one has
\beqa
\eta=\pi_*(c_2(V)).
\eeqa
Now, let us turn to the case that $n$ is odd! \\ \\
\underline{$n$ odd}\\ \\
Let us first specify our unstable bundle $V$. Actually we have $n$ 
different choices to do this which depend on the choice of the integer $k$
in the range $1\le k \le n$. We will choose 
$k=\frac{n+1}{2}$ and the line bundles ${\cal M}={\cal S}^{\frac{-n+1}{2}}$ 
and ${\cal M}'={\cal S}^{\frac{n+1}{2}}\otimes {\L}^{-1}$ which will
presently be shown to be  
appropriate in order to be compared with results 
obtained from spectral covers for $V$. Therefore
we can write for our unstable bundle $V$
\beqa
V=W_k\otimes {\cal S}^{-n+k}\oplus W^*_{n-k}\otimes {\cal S}^k\otimes {\L}^{-1}
\eeqa
which has trivial determinant. Using the above relation for the total 
Chern-class of $V$ and setting again $\sigma=c_1({\cal O}(\sigma))$ 
and ${\sigma}^2= -\sigma c_1({\L})$, and
\beqa
\eta= n c_1({\cal S})
\eeqa
we will find for the characteristic classes of $V$ 
\beqa
c_1(V)&=&0\\
c_2(V)&=&\eta\sigma-\frac{1}{24}c_1(\L)^2(n^3-n)-
       \frac{n}{8}\eta(\eta-nc_1(\L))+\frac{1}{8n}\eta(\eta-nc_1(\L))\label{1}
\\
c_3(V)&=&\frac{1}{n}\sigma\eta(\eta-nc_1(\L))
\eeqa
So, we are restricted to bundles $V$ with $\eta=\pi_*(c_2(V))$ 
divisible by $n$. 

Now, the integration of $c_3$ over $Z$ can be accomplished by first integrating
over the fibers of $Z\rightarrow B$ and then integrating over the base. 
Further, using the fact that the section $\sigma$ intersects the fiber $F$
in one point $\sigma\cdot F=1$ and $r=3\sigma$ where $r$ is the cohomology
class dual to the vanishing of the section of the line bundle ${\cal O}(1)$
(defined on the total space of the Weierstrass model of $Z$), we get
\beqa
\int_Z c_3(V)=\int_B \frac{1}{n}\eta(\eta-nc_1(\L)).
\label{pa}
\eeqa
Further recall, that the net number of generations $N_{gen}$ of chiral fermions
in the observable sector of the gauge group is given by (c.f. [\ref{GSW}]) 
\beqa
N_{gen}=\frac{1}{2}\mid{\int_{Z} c_3(V)}\mid
\eeqa
which reflects the fact that massless fermions in four dimensions correspond
to zero modes of the Dirac operator on the Calabi-Yau manifold $Z$. 
So we get
\beqa
N_{gen}=n_{gen}-{\bar n}_{gen}=\int_B \frac{1}{2n}\eta(\eta-nc_1(\L))
\eeqa
and thus, if we fix the elliptic manifold $Z$, so that the section $\sigma$
and $c_1(\L)$ are fixed, $N_{gen}$ is uniquely determined for a choice of 
$\eta$. Here we have to note that in order to separately determine 
the number of generations $n_{gen}$ respectively, 
antigenerations ${\bar n}_{gen}$ instead
of just their difference, we have to compute in addition the dimension of
$H^1(Z,V)$. This can be done by using the Leray spectral sequence.\\
\\ 
{\bf{Remarks:}} Using $N_{gen}$ 
we can write for the total number of $N=1$ chiral matter 
multiplets which are charged under the unbroken gauge group 
$C^c_{het}=N_{gen}+2{\bar n}_{gen}$.
Recall that the number of $N=1$ neutral chiral (resp. antichiral)
multiplets $C_{het}$ is given by (c.f.[\ref{ACL}])
$C_{het}=h^{21}(Z)+h^{11}(Z)+m_{bun}$
with $m_{bun}$ denoting the number of bundle moduli, i.e. the
dimension of $H^1(Z,ad(V))$.
Note, in the former case of $V=SU(n)$ with $n$ even, which was
discusse in [\ref{FMW}], we have $C^c_{het}=2{\bar n}_{gen}$.

\subsection{Comparison With The Spectral Cover Construction}

In order to compare our results with those obtained from the spectral 
cover construction let us review some facts of the setup [\ref{FMW}]. 

The spectral cover $C$ is given by the vanishing of a section $s$ of
${\cal O}(\sigma)^n\otimes {\cal M}$ with $\M$ being an arbitrary
line bundle over $B$ 
of $c_1(\M)=\eta'$. The locus $s=0$ for the section is given for $n$ even by 
$s=a_0z^n+a_2z^{n-2}x+a_3z^{n-3}y+...+a_nx^{n/2}$ 
(resp. the last term is $x^{(n-3)/2}y$ for n odd)\footnote {here 
$a_{r}\in\Gamma(B,\M\otimes\L^{-r})$, $a_0$ is a section of $\M$ and $x$, $y$ 
sections of $\L^2$ resp. $\L^3$ in the Weierstrass model (c.f.[\ref{FMW}])}.
Further, one has a twisting
line bundle ${\cal N}$ over $C$ so that we have the vector bundle 
$V=\pi_{2*}({\cal N}\otimes {\cal P}_B)$ over $Z$ with $\pi_2$ being the 
projection of $C\times_B Z$ to the second factor. Now, since the 
Poincare line bundle ${\cal P_B}$ becomes trivial when restricted to $\sigma$
and with Grothendieck-Riemann-Roch for the projection 
$C\rightarrow B$, we get
\beqa
\pi_*(e^{c_1({\cal N})}Td(C))=ch(V)Td(B)
\eeqa
and with the condition $c_1(V)=0$ one has 
\beqa
c_1({\cal N})=-\frac{1}{2}(c_1(C)-\pi_*c_1(B))+\gamma
=\frac{1}{2}(n\sigma+\eta'+c_1(\L))+\gamma
\eeqa
here $\gamma\in H^{1,1}(C,{\bf Z})$ with $\pi_*\gamma=0\in H^{1,1}(B,{\bf Z})$.
In particular if one denotes by $K_B$ and $K_C$ the canonical bundles of $B$ 
and $C$ then one has (c.f.[\ref{FMW}])
\beqa
{\cal N}=K_C^{1/2}\otimes K_B^{-1/2}\otimes ({\cal O}(\sigma)^n\otimes\M^{-1}
\otimes \L^n)^{\lambda}
\eeqa
from which one learns that $\gamma=\lambda(n\sigma-\eta'+nc_1(\L))$. 

The 
second Chern class of $V$ is given by
\beqa
c_2(V)=\eta'\sigma-\frac{1}{24}c_1(\L)^2(n^3-n)-\frac{n}{8}\eta'(\eta'
-nc_1(\L))-\frac{1}{2}\pi_*(\gamma^2)
\label{om}
\eeqa
where the last term reflects the fact that one can twist with a line bundle
${\cal N}$ on the spectral cover, one has
\beqa
\pi_*(\gamma^2)=-\lambda^2n\eta'(\eta'-nc_1(\L)).
\eeqa
Now let us compare!

In case that $n$ is even it was shown [\ref{FMW}] that to achieve 
$\tau$ invariance in the spectral cover approach for $V$, one must define
${\cal N}$ in the above sense with $\gamma=0$, i.e. $\lambda=0$. In 
particular it was shown the existence of an isomorphism ${\cal N}^2=K_C\otimes
K_B^{-1}$. Further it was shown that there are the same mod two coditions 
for $\eta'$ and $n$ in the spectral cover approach for $\gamma=0$ and in 
particular that also $\eta'=\pi_*(c_2(V))$ and therefore one is lead to the
identification $\eta'=\eta$. 

Also in case that $n$ is even and $\lambda=\frac{1}{2}$ 
the last two terms in (\ref{om}) combine (c.f.[\ref{FMW}],[\ref{CD}]) the only
general elements of $H^{1,1}(C,{\bf Z})$ are 
$\sigma|_C$ and $\pi^* \beta$ (for $\beta \in H^{1,1}(B,{\bf Z})$), 
which have because of $C=n\sigma + \pi ^*\pi_* c_2V$
the relation $\pi_* (\sigma|C)=\pi_* \sigma(n\sigma+\pi^*\eta)=
\pi_* \sigma(-nc_1+\pi^*\eta)=\eta-nc_1$;
so $\gamma=\lambda (n\sigma-\pi^*(\eta-nc_1))$ (with $\lambda$ 
possibly half-integral) and $\pi_*(\gamma^2)=-\lambda^2 n\eta(\eta-nc_1)$;
so for $\lambda=1/2$ the term would disappear.

Now if $n$ is odd, we can identify the last term in ({\ref{1}})
with $\frac{\pi_*(\gamma^2)}{2}$ if we choose $\lambda=\frac{1}{2n}$, i.e. 
the parabolic approach for $n$ odd agrees with the spectral cover approach
if we choose the twisting line bundle ${\cal N}$ appropriate on the 
spectral cover. Furthermore, if we use 
\beqa
c_1({\cal N})=\frac{1}{2}(n\sigma+\eta'+c_1(\L))+\gamma=\frac{(n+1)}{2}\sigma
+c_1(\L)+\frac{(n-1)}{2} \frac{\eta'}{n}
\eeqa
which is well defined for $n$ odd and since we can choose 
${\cal M}={\cal S}^n$ we are left with ${\eta}'=n c_1({\cal S})$ and so 
with $\lambda=\frac{1}{2n}$ we have the same conditions 
for $\eta'$ and $n$ as we had in the parabolic approach.\\
\\ 
{\bf{Discussion:}} We have constructed a class of 
$SU(n)$ vector bundles, with $n$ odd, 
in the parabolic bundle construction, which have a 
$\eta\equiv 0 ({\rm{mod}}\ \ n)$ condition,
in contrast to the bundles, which have $n$ even and a
$\eta\equiv c_1(\L) ({\rm{mod}}\ \ 2$)) condition.  
For $n$ even, the bundles in the parabolic construction are restricted 
to the $\tau$-invariant bundles
in the spectral cover construction, given at $\lambda=0$. Our bundles have no 
$\tau$-invariance and being restricted to bundles of
$\lambda=\frac{1}{2n}$, in the  
spectral cover construction.
\subsection{A Comment On Some Discrete Data}
Let us now recall, it was shown [\ref{flux}] that one 
has as quantization law for the
four-flux $G=\frac{1}{2\pi}dC$ the modified integrality condition
$G=\frac{c_2}{2}+\alpha$ with $\alpha\in H^4(X,{\bf Z})$. Furthermore, it was
shown [\ref{CD}] that $\alpha$ is further restricted by 
$\int \alpha^2+\alpha c_2 \le -120$ in order to keep the wanted amount 
of supersymmetry in a consistent compactification. It has also been shown 
[\ref{DM}] 
that the appearance of the four-flux modifies the number of space-time
filling threebranes $n_3$ which are necessary for tadpole 
cancellation in $F$-theory
[\ref{SVW}]; one has
\beqa
n_3=\frac{\chi(X^4)}{24}-\frac{1}{2}G^2. 
\eeqa
Now a recent paper showed [\ref{CD}] that on the heterotic side
the additional degree of freedom, coming from possible twists of the line
bundle ${\cal N}$ on the spectral cover modifies the number of heterotic
fivebranes 
\beqa
n_5(\gamma)=n_5(\gamma=0)+\frac{1}{2}\pi_*(\gamma^2).
\eeqa
Moreover, [\ref{CD}] showed that the heterotic "twists" correspond
to the appearance of a non-trivial four-flux on the $F$-theory side
\beqa
\pi_*(\gamma^2)=-G^2
\eeqa
and therefore rounded off the picture that the number of heterotic fivebranes 
matches the number of $F$-theory threebranes which was established in 
[\ref{FMW}] for $E_8$ and in [\ref{AC}] for $SU(n)$ vector bundle.

Now, using the identification of $\frac{1}{8n}\eta(\eta-nc_1(\L))$ with
$\frac{1}{2}\pi_*(\gamma^2)$ at $\lambda=\frac{1}{2n}$, denoting this by 
$\frac{1}{2}\pi_*(\gamma^2)|_{\lambda=1/2n}=\frac{1}{2}\pi_*(\gamma^2_n)$, we 
can write the third Chern class at $\lambda=\frac{1}{2n}$
\beqa
\int_{B}c_3(V)=4\int_B \pi_*(\gamma^2_n).
\label{pi}
\eeqa
and therefore see that $N_{gen}$ is related to the appearance of the 
four-flux in F-theory.\footnote{ At this point I am very grateful to
G. Curio for many inspiring discussions and inparticular for pointing out  
such a relation to me.}\\ \\   
{\bf{Note added:}} In a recent paper [\ref{CC}], 
the computation of $c_3(V)$ was performed in the spectral cover approach,
it is given by (in the $l=0$ sector c.f. [\ref{CC}])
\beqa
\int_B c_3(V)=2\int_B \lambda \eta(\eta-nc_1).
\eeqa
For $\lambda=\frac{1}{2n}$,
this is in nice agreement with our computation of $c_3(V)$ in the 
parabolic approach (\ref{pa}). Further, (\ref{pi}) implies 
$c_3(V)\sim \lambda^2n \eta(\eta-nc_1)$ away from $\lambda=\frac{1}{2n}$
but actually one has $c_3(V)\sim \lambda\eta(\eta-nc_1)$.\\
%------------------------------------------------------------------------------
{\bf Acknowledgements:} I would like to thank G. Curio and D. L\"ust  
for discussions.
%------------------------------------------------------------------------------

\center{\bf References}
\begin{enumerate}
%------------------------------------------------------------------------------
\item
\label{FMW}
R. Friedman, J. Morgan and E. Witten, {\it Vector Bundles and 
F-Theory}, Commun. Math. Phys. {\bf 187} (1997) 679, hep-th/9701162.

\item
\label{V} 
C. Vafa, {\em Evidence for F-Theory}, Nucl. Phys. {\bf B469} (1996) 493,
hep-th/9602022.

\item
\label{MV} 
D. R. Morrison and C. Vafa, {\em Compactification of F-theory on 
Calabi-Yau
Threefolds I,II}, Nucl. Phys. {\bf B 473} (1996) 74, hep-th/9602114;
Nucl. Phys. {\bf B 476} (1996) 437, hep-th/9603161.

\item
\label{BJPS}
M. Bershadsky, A. Johansen, T. Pantev and V. Sadov, {\it On Four-Dimensional 
Compactifications of F-Theory}, Nucl. Phys. {\bf B505} (1997), 165-201, 
hep-th/9701165.

\item
\label{SVW}
S. Sethi, C. Vafa and E. Witten, {\it Constraints on Low-Dimensional String
Compactifications}, Nucl. Phys. {\bf B 480} (1996) 213, hep-th/9606122.

\item
\label{AC}
B. Andreas and G. Curio, {\it Three-branes and five-branes in $N=1$ dual string
pairs}, Phys. Lett. {\bf B417} (1998) 41-44, hep-th/9706093.

\item
\label{CD}
G. Curio and R. Donagi, {\it Moduli in N=1 heterotic/F-theory duality},
hep-th/9801057.

\item
\label{D}
R. Y. Donagi, {\it ICMP lecture on Heterotic/F-theory duality}, hep-th/9802093.

\item
\label{D1}
R. Y. Donagi, {\it Principal bundles on elliptic fibrations}, Asian J. Math.
{\bf 1} (1997), 214-223, alg-geom/9702002.

\item
\label{ACL}
B. Andreas, G. Curio and D. L\"ust, {\it $N=1$ Dual String Pairs and their
Massless Spectra}, Nucl. Phys. {\bf B507} (1997) 175, hep-th/9705174.

\item
\label{AM}
P. S. Aspinwall and D. R. Morrison, {\it Point like Instantons on K3
Orbifolds}, Nucl. Phys. {\bf B503} (1997) 533-564, hep-th/9705104.

\item
\label{BCGJL}
M. Bershadsky, T. M. Chiang, B. R. Greene, A. Johansen and C. I. Lazaroiu,
{\it F-theory and Linear Sigma Models}, hep-th/9712023.

\item
\label{KV}
S. Katz and C. Vafa, {\it Matter from Geometry}, Nucl. Phys. {\bf 497} (1997)
146-154, hep-th/9606086.

\item
\label{geom}
S. Katz, A. Klemm and C. Vafa, {Geometric Engineering of Quantum Field
Theories}, Nucl. Phys. {\bf B497} (1997) 173-195, hep-th/9609239.

\item
\label{KV2}
S. Katz and C. Vafa, {Geometric Engineering of N=1 Quantum Field
Theories}, Nucl. Phys. {\bf B497} (1997) 196-204, hep-th/9611090.

\item
\label{BIKMSV} 
M. Bershadsky, K. Intriligator, S. Kachru, D. Morrison,
V. Sadov and C. Vafa, {\it Geometric Singularities and Enhanced Gauge
Symmetries}, Nucl. Phys. {\bf B 481} (1996) 215, hep-th/9605200.

\item
\label{BJPSV} 
M. Bershadsky, A. Johansen, T. Pantev, V. Sadov and C. Vafa,
{\em $F$-theory, Geometric Engineering and $N=1$ Dualities},
Nucl. Phys. {\bf B505} (1997) 153-164, hep-th/9612052.

\item
\label{ler}
W. Lerche, {\it Fayet-Iliopoulos Potentials from Four-Folds}, hep-th/9709146.

\item
\label{sup1}
E. Witten, {\it Nonperturbative Superpotentials in String Theory},
Nucl. Phys. {\bf B 474} (1996) 343, hep-th/9604030.

\item
\label{sup2}
R. Donagi, A. Grassi and E. Witten, {\em A Non-perturbative Superpotential
with $E_8$ Symmetry}, Mod. Phys. Lett. {\bf A 11} (1996) 2199.

\item
\label{sup3}
G. Curio and D. L\"ust, {\em A Class of $N=1$ Dual String Pairs
and its Modular Superpotential}, Int. Journ.
of Mod. Phys. {\bf A12} (1997) 5847-5866, hep-th/9703007.

\item
\label{sup4}
P. Mayr, {\it Mirror Symmetry, N=1 Superpotentials and Tensionless Strings
on Calabi-Yau Four-Folds}, Nucl. Phys. {\bf B494} (1997) 489-545,
hep-th/9610162.

\item
\label{D2}
R. Y. Donagi, {\it Spectral covers, in: Current topics in complex algebraic
geometry}, MSRI pub. {\bf 28} (1992), 65-86, alg-geom/9505009.

\item
\label{FMW2}
R. Friedman, J. Morgan and E. Witten, {\it Principal $G$-Bundles Over Elliptic
Curves}, alg-geom/9707004.

\item
\label{FMW3}
R. Friedman, J. Morgan and E. Witten, {\it Vector Bundle Over Elliptic 
Fibrations}, alg-geom/9709029.

\item
\label{GSW}
M. Green, J. Schwarz and E. Witten, {\it Superstring Theory}, v.2,
Cambridge University Press (1987).

\item
\label{DM}
K. Dasgupta and S. Mukhi, {\it A Note on Low-Dimensional String 
Compactifications}, Phys. Lett. {\bf B398} (1997) 285, hep-th/9612188.

\item
\label{flux}
E. Witten, {\it On Flux Quantization in M Theory and the Effective Action},
J. Geom. Phys. {\bf 22} (1997) 1, hep-th/9609122.

\item
\label{CC}
G. Curio, {\it Chiral matter and transitions in heterotic string models},
hep-th/9803224.

\end{enumerate}
\end{document}